# Probing top quark decay into light stop in the supersymmetric standard model at the upgraded Tevatron

M. Hosch[a], R. J. Oakes[b], K. Whisnant[a], Jin Min Yang[b,1], Bing-Lin Young[a] and X. Zhang[c]

[a] *Department of Physics and Astronomy, Iowa State University,*

*Ames, Iowa 50011, USA*

[b] *Department of Physics and Astronomy, Northwestern University,*

*Evanston, Illinois 60208, USA*

[c] *CCAST (World Laboratory), P.O. Box 8730, Beijing 100080, and*

*Institute of High Energy Physics, Academia Sinica, Beijing 100039, China*

ABSTRACT

We investigate the possibility of observing the exotic decay mode of the top quark into the lightest stop ($\tilde{t}_1$) and neutralino ($\tilde{\chi}_1^0$) in the minimal supersymmetric standard model with R-parity at the upgraded Tevatron. First we determine the allowed range for the branching fraction $B(t \to \tilde{t}_1 \tilde{\chi}_1^0)$ in the region of parameter space allowed by the $R_b$ data and the CDF $ee\gamma\gamma + \not{E}_T$ event, and then consider all possible backgrounds and investigate the possibility of observing this final state at the Tevatron. We find that this final state is unobservable at Run 1. However, Run 2 can provide significant information on this new decay mode of the top quark: either discover it, or establish a strong constraint on the masses of $\tilde{t}_1$ and $\tilde{\chi}_1^0$ given approximately by $M_{\tilde{\chi}_1^0} > M_{\tilde{t}_1} - 6$ GeV.

---

[1]On leave from Physics Department, Henan Normal University, China

## 1. Introduction

Because of its large mass, the top quark has the potential to be a sensitive probe for new physics. In strongly interacting theories, such as top condensation and extended technicolor, the top quark plays an essential role in the electroweak symmetry breaking and in the understanding of flavor physics. In weakly interacting theories, such as supersymmetry (SUSY) [1], the heavy top quark provides a solution to the electroweak symmetry breaking and makes it possible that the top quark may decay into its lightest superpartner [2] ($\tilde{t}_1$) plus the lightest neutralino ($\tilde{\chi}_1^0$). Assuming that $\tilde{t}_1$ decays dominantly into $c\tilde{\chi}_1^0$, this SUSY decay mode of the top quark will give rise to a new final state in $t\bar{t}$ production at the Fermilab Tevatron, $t\bar{t} \to Wbc\tilde{\chi}_1^0\tilde{\chi}_1^0$.

A careful study of this final state is well motivated since the presently allowed parameter space [3][4] of the minimal supersymmetric standard model (MSSM), implied by the $ee\gamma\gamma + \not{E}_T$ event at CDF [5] and the LEP $R_b$ data, will allow the top quark to have this new decay mode $t \to \tilde{t}_1\tilde{\chi}_1^0$. It is also implied in this scenario that $M_{\tilde{t}_1} < M_{\tilde{\chi}_1^\pm}$ [3,4], where $\tilde{\chi}_1^\pm$ is the lightest chargino, so the decay $\tilde{t}_1 \to b\tilde{\chi}_1^+$ is not allowed. In the case that the $\tilde{t}_1$ is lighter than the next-to-lightest neutralino $\tilde{\chi}_2^0$, $\tilde{l}^\pm$ and $\tilde{\nu}$, the dominant decay of $\tilde{t}_1$ is $\tilde{t}_1 \to c\tilde{\chi}_1^0$ via one loop processes [6], with a branching fraction of almost 100% [3]. Therefore searching for this final state may be a powerful tool for probing SUSY at FNAL.

The possibility for detecting the $Wbc\tilde{\chi}_1^0\tilde{\chi}_1^0$ final state in $t\bar{t}$ production was first discussed in Ref. [7] where the focus was mainly on the background $t\bar{t} \to W^-W^+b\bar{b}$. In this article, in the framework of the MSSM with the lightest neutralino being the LSP, we will present a detailed analysis including all the possible backgrounds. In particular, we first determine the allowed range for $B(t \to \tilde{t}_1\tilde{\chi}_1^0)$ in the region of parameter space allowed by the $R_b$ data and the $ee\gamma\gamma + \not{E}_T$ event at CDF, and then show what additional constraints can be imposed on the allowed parameter space if this final state is not observed at the Tevatron. We find a lower bound of 0.07 for $B(t \to \tilde{t}_1\tilde{\chi}_1^0)$ in the presently allowed parameter space [3][4], and, as a result, we find that Run 2 can either discover this new decay mode or provide an additional strong constraint given approximately by $M_{\tilde{\chi}_1^0} > M_{\tilde{t}_1} - 6$ GeV. However, our results show that even if $B(t \to \tilde{t}_1\tilde{\chi}_1^0)$ is as large as 0.5, it is unobservable at Run 1.

This paper is organized as follows. In Sec. 2 we examine the range of values for $B(t \to \tilde{t}_1\tilde{\chi}_1^0)$ in the region of parameter space allowed by the $R_b$ data and the $ee\gamma\gamma + \not{E}_T$ event.

---

[2] Electroweak baryogenesis in SUSY requires a light stop to have a strong first order phase transition [2].
[3] The four-body decay mode $\tilde{t}_1 \to b\tilde{\chi}_1^0 f_1\bar{f}_2$ is kinematically suppressed by both $\tilde{\chi}_1^\pm-$ and $W^\pm-$ propagators and thus its partial width is negligibly small.

In Sec. 3 we examine all possible backgrounds and investigate the possibility of observing $t\bar{t} \to Wbc\tilde{\chi}_1^0\tilde{\chi}_1^0$ at the Tevatron. And finally in Sec. 4 we present a summary.

## 2. Bounds for $B(t \to \tilde{t}_1\tilde{\chi}_1^0)$

It was found that in order to explain all of the presently available low energy data the lightest mass eigenstate ($\tilde{t}_1$) of the stop squarks is likely to be the right-stop ($\tilde{t}_R$) with mass of the order of $M_W$ [4]. So we assume $\tilde{t}_1 = \tilde{t}_R$ in our analyses. The interaction Lagrangian of top ($t$), stop($\tilde{t}_1$) and neutralino ($\tilde{\chi}_1^0$) is given by [8]

$$\mathcal{L}_{t\tilde{t}_1\tilde{\chi}_1^0} = -\sqrt{2}\bar{t}(AP_L + BP_R)\tilde{\chi}_1^0\tilde{t}_1 + H.c., \tag{1}$$

where $P_{L,R} = (1 \mp \gamma_5)/2$ and

$$A = -\frac{2}{3}e(N_{11}C_W + N_{12}S_W) + \frac{2}{3}\frac{gS_W^2}{C_W}(N_{12}C_W - N_{11}S_W), \tag{2}$$

$$B = -\frac{gM_t N_{14}}{2M_W \sin\beta}. \tag{3}$$

Here $S_W \equiv \sin\theta_W$, $C_W \equiv \cos\theta_W$, and $N_{ij}$ are the elements of the $4 \times 4$ matrix $N$ which diagonalizes the neutralino mass matrix [1]. The decay $t \to \tilde{t}_1\tilde{\chi}_1^0$ has been calculated to one-loop level in Refs. [9] and [10]. Here we neglect the loop corrections, which are only on the order of 10%; the partial width is given at tree level by

$$\Gamma(t \to \tilde{t}_1\tilde{\chi}_1^0) = \frac{1}{16\pi M_t^3}\lambda^{1/2}(M_t^2, M_{\tilde{\chi}_1^0}^2, M_{\tilde{t}_1}^2)\left[(|A|^2 + |B|^2)(M_t^2 + M_{\tilde{\chi}_1^0}^2 - M_{\tilde{t}_1}^2)\right.$$
$$\left. +4Re(A^*B)M_t M_{\tilde{\chi}_1^0}\right], \tag{4}$$

where $\lambda(x, y, z) = (x - y - z)^2 - 4yz$.

The parameters involved in $\Gamma(t \to \tilde{t}_1\tilde{\chi}_1^0)$ are:

$$M_{\tilde{t}_1}, M_2, M_1, \mu, \tan\beta, \tag{5}$$

where $M_2$ and $M_1$ are gaugino masses corresponding to $SU(2)$ and $U(1)$, $\mu$ is the coefficient of the $H_1H_2$ mixing term in the superpotential, and $\tan\beta = v_2/v_1$ is the ratio of the vacuum expectation values of the two Higgs doublets. If the $ee\gamma\gamma + \not{E}_T$ event is due to $\tilde{e}_L$ pair production (depending on the scenario, the selectrons decay into either $e\tilde{\chi}_2^0$ or $e\tilde{\chi}_1^0$, followed by $\tilde{\chi}_2^0 \to \tilde{\chi}_1^0 \gamma$ or $\tilde{\chi}_1^0 \to \tilde{G}\gamma$, respectively, where $\tilde{G}$ is the gravitino), the region of the parameter space allowed by both the $ee\gamma\gamma + \not{E}_T$ event and the $R_b$ data are given as [3]

$$50 \leq M_1 \leq 92 \text{ GeV}, \quad 50 \leq M_2 \leq 105 \text{ GeV},$$
$$0.75 \leq M_2/M_1 \leq 1.6, \quad -65 \leq \mu \leq -35 \text{ GeV},$$
$$0.5 \leq |\mu|/M_1 \leq 0.95, \quad 1 \leq \tan\beta \leq 3,$$
$$33 \leq M_{\tilde{\chi}_1^0} \leq 55 \text{ GeV}, \quad 45 \leq M_{\tilde{t}_1} \leq 80 \text{ GeV}. \quad (6)$$

If the $ee\gamma\gamma + \not{E}_T$ event is due to $\tilde{e}_R$ pair production, the allowed region is [3]

$$60 \leq M_1 \leq 85 \text{ GeV}, \quad 40 \leq M_2 \leq 85 \text{ GeV},$$
$$0.6 \leq M_2/M_1 \leq 1.15, \quad -60 \leq \mu \leq -35 \text{ GeV},$$
$$0.5 \leq |\mu|/M_1 \leq 0.8, \quad 1 \leq \tan\beta \leq 2.2,$$
$$32 \leq M_{\tilde{\chi}_1^0} \leq 50 \text{ GeV}, \quad 45 \leq M_{\tilde{t}_1} \leq 80 \text{ GeV}. \quad (7)$$

In the region of Eq.(6) we obtain

$$0.07 \leq B(t \to \tilde{t}_1 \tilde{\chi}_1^0) \leq 0.50, \quad (8)$$

and in the region of Eq.(7),

$$0.10 \leq B(t \to \tilde{t}_1 \tilde{\chi}_1^0) \leq 0.50. \quad (9)$$

So, if the $ee\gamma\gamma + \not{E}_T$ event is explained by the MSSM with the lightest neutralino being the LSP, the exotic decay $t \to \tilde{t}_1 \tilde{\chi}_1^0$ must occur at a branching ratio larger than 0.07.

Upper bounds for this exotic decay of the top quark can also be derived from the available data at FNAL. Currently, the FNAL top quark pair production counting rate is interpreted as a measurement of $\sigma(t\bar{t}) \times B^2(t \to bW)$. Since the final states $t\bar{t} \to Wb\bar{c}\tilde{\chi}_1^0\tilde{\chi}_1^0$ and $t\bar{t} \to c\tilde{\chi}_1^0\tilde{\chi}_1^0\bar{c}\tilde{\chi}_1^0\tilde{\chi}_1^0$ do not have enough leptons or jets to be included in the dileptonic, leptonic or hadronic event samples, they are invisible to the current counting experiments at FNAL. So the quantity $[1 - B(t \to \tilde{t}_1 \tilde{\chi}_1^0)]^2$, which gives the fraction of events in which both the $t$ and $\bar{t}$ decay normally [4], should lie within the measured range of $\sigma[t\bar{t}]_{\text{exp}}/\sigma[t\bar{t}]_{\text{QCD}}$. Note that in our analyses we neglected the SUSY effects [13][14] in $t\bar{t}$ production and thus the theoretical value of $\sigma[t\bar{t}]$ is given by the SM value $\sigma[t\bar{t}]_{\text{QCD}}$. The production cross section measured by CDF with an integrated luminosity of 110 pb$^{-1}$ is $\sigma[t\bar{t}]_{\text{exp}} = 8.5^{+4.4}_{-3.4}, 6.8^{+2.3}_{-1.8}, 10.7^{+7.6}_{-4.4}$ pb in the dilepton, lepton+jets and all-hadronic channels, respectively [15]. The SM expectation for top mass of 175 GeV is $\sigma[t\bar{t}]_{\text{QCD}} = 5.5^{+0.1}_{-0.4}$ pb [16]. By comparing $\sigma[t\bar{t}]_{\text{exp}}$ from each channel with $\sigma[t\bar{t}]_{\text{QCD}}[1 - B(t \to \tilde{t}_1 \tilde{\chi}_1^0)]^2$, we find that the $2\sigma$ upper bounds on $B(t \to \tilde{t}_1 \tilde{\chi}_1^0)$ for the various channels are given by

$$B(t \to \tilde{t}_1 \tilde{\chi}_1^0) \leq \begin{cases} 0.44 & \text{dilepton channel} \\ 0.23 & \text{lepton + jets channel} \\ 0.41 & \text{all} - \text{hadronic channel}. \end{cases} \quad (10)$$

---

[4]Here we assume that the only exotic decay mode of top quark in R-parity conserving MSSM is $t \to \tilde{t}_1 \tilde{\chi}_1^0$. If charged Higgs is light enough, $t \to H^+ b$ is also possible; its phenomenological implications at Tevatron have been studied [11]. The FCNC decays $t \to cZ, c\gamma, cg, ch$ are negligibly small in R-parity conserving MSSM [12].

Here the upper bound from lepton+jets channel is comparable to the upper bound of 0.25 [7] obtained by a global fit to the available data. If the possible enhancement of $t\bar{t}$ production cross section from gluino pair production is taken into account, the upper bound for $B(t \to \tilde{t}_1 \tilde{\chi}_1^0)$ can be relaxed to 0.5 [14].

## 3. Observing $t\bar{t} \to Wbc\tilde{\chi}_1^0\tilde{\chi}_1^0$ at the Tevatron

Under the assumption that the top (or anti-top) decays via the normal weak interactions to $Wb$, the anti-top (top) decays to $\tilde{t}_1\tilde{\chi}_1^0$, and the light stop decays to $c\tilde{\chi}_1^0$, then the final state of interest is $Wbc\tilde{\chi}_1^0\tilde{\chi}_1^0$. Due to the large QCD backgrounds, it is very difficult to search for the signal from the hadronic decays of $W$ at the Tevatron. We therefore look for events with the leptonic decay of the $W$. Thus, the signature of this process is an energetic charged lepton, one $b$-quark jet, one light $c$-quark jet, plus missing $E_T$ from the neutrino and the unobservable $\chi_1^{0\prime}$s. We assumed silicon vertex tagging of the $b$-quark jet with 50% efficiency and the probability of 0.4% for a light quark jet to be mis-identified as a $b$-jet. The potential SM backgrounds are:

(1) $bq(\bar{q}) \to tq'(\bar{q}')$;

(2) $q\bar{q}' \to W^* \to t\bar{b}$;

(3) $Wb\bar{b}$;

(4) $Wjj$;

(5) $t\bar{t} \to W^-W^+b\bar{b}$;

(6) $gb \to tW$;

(7) $qg \to q't\bar{b}$.

The quark-gluon process (7) can occur with a W-boson intermediate state in either the t-channel or the s-channel. We found backgrounds (6) and (7) to be negligible since they have an extra jet, and can mimic our signal (before b-tagging) only if a jet is missed in the detector. The background process (5) can mimic our signal if both $W$'s decay leptonically and one charged lepton is not detected, which we assumed to occur if the lepton pseudo-rapidity and transverse momentum satisfy $\eta(l) > 3$ and $p_T(l) < 10$ GeV, respectively.

To simulate the detector acceptance, we made a series of basic cuts on the transverse momentum ($p_T$), the pseudo-rapidity ($\eta$), and the separation in the azimuthal angle-pseudo-rapidity plane ( $\Delta R = \sqrt{(\Delta\phi)^2 + (\Delta\eta)^2}$ ) between a jet and a lepton or between two jets. These cuts are chosen to be

$$p_T^l,\ p_T^{\text{jet}},\ p_T^{\text{miss}} \geq 20 \text{ GeV}, \tag{11}$$

$$\eta_{\text{jet}},\ \eta_l \leq 2.5, \tag{12}$$

$$\Delta R_{jj},\ \Delta R_{jl} \geq 0.5. \tag{13}$$

Further simulation of detector effects is made by assuming a Gaussian smearing of the energy of the final state particles, given by:

$$\Delta E/E = 30\%/\sqrt{E} \oplus 1\%, \text{ for leptons}, \tag{14}$$

$$= 80\%/\sqrt{E} \oplus 5\%, \text{ for hadrons}, \tag{15}$$

where $\oplus$ indicates that the energy dependent and independent terms are added in quadrature and $E$ is in GeV.

In order to substantially reduce the background, we apply a cut on the transverse mass defined by $m_T = \sqrt{(P_T^l + P_T^{\text{miss}})^2 - (\vec{P}_T^l + \vec{P}_T^{\text{miss}})^2}$. Without smearing, $m_T$ is always less than $M_W$ (and peaks just below $M_W$) if the only missing energy comes from a neutrino from W decay, which is the case for most of the background events (single top, Wbb, Wjj). Smearing pushes some of this above $M_W$. For the signal $m_T$ is spread about equally above and below $M_W$, due to the extra missing energy of the neutralinos. Therefore we also require

$$m_T > 90 GeV. \tag{16}$$

The results with different cuts are shown in Table 1. For convenience the numerical results shown are obtained without including the appropriate branching ratios; the actual cross sections are found by multiplying the given values by the branching fraction factors $x = B(t \to c\tilde{\chi}_1^0\tilde{\chi}_1^0)$ and $1 - x = B(t \to bW)$. The products of the appropriate branching fractions in each case are given in the last column of Table 1. In our numerical evaluation, we assumed $M_t = 175$ GeV, $\sqrt{s} = 1.8$ TeV and an integrated luminosity of 0.1 fb$^{-1}$ for Run 1, and $\sqrt{s} = 2$ TeV and an integrated luminosity of 10 fb$^{-1}$ for Run 2.

At Run 1, with the basic and $m_T$ cuts the number of background events is always less than 1, and the number of signal events is always less than 9 for any value of $x = B(t \to \tilde{t}_1 \tilde{\chi}_1^0)$ allowed by Eq.(8) or (9). Thus the signal is unobservable at Run 1 under the criteria $S \geq 3\sqrt{B+S}$, which corresponds to the 95% confidence level (C.L.) if we assume Poisson statistics. The $m_T$ cut hurts the signal, but, as we pointed out above, it reduces the background much more than the signal. Even when the $m_T$ cut is relaxed, this final state is unobservable at Run 1.

At Run 2 this signal is observable even for quite small values of $B(t \to \tilde{t}_1 \tilde{\chi}_1^0)$. In Fig. 1 we show $B(t \to \tilde{t}_1 \tilde{\chi}_1^0)$ versus $M_{\tilde{\chi}_1^0}$ for the signal to be observable at 95% C.L. The region above each curve is the corresponding observable region. From this figure we can see that for $B(t \to \tilde{t}_1 \tilde{\chi}_1^0) > 0.07$ (the minimum value allowed by Eqs. 8 and 9), Run 2 can detect this signal at 95% C.L. if (approximately) $M_{\tilde{\chi}_1^0} \leq M_{\tilde{t}_1} - 5$ GeV. This means that if this signal is not observed at Run 2, an additional constraint, given approximately by

$$M_{\tilde{\chi}_1^0} > M_{\tilde{t}_1} - 5 \text{ GeV}, \tag{17}$$

can be placed on the region of Eq.(8) allowed by the $R_b$ data and the $ee\gamma\gamma + \not{E}_T$ event at 95% C.L.

In Fig. 2 we present $B(t \to \tilde{t}_1 \tilde{\chi}_1^0)$ versus $M_{\tilde{\chi}_1^0}$ for the signal to be observable under the stricter discovery criteria $S \geq 5\sqrt{B}$. In this case we see that if this signal is not observed at Run 2 the additional constraint of approximately

$$M_{\tilde{\chi}_1^0} > M_{\tilde{t}_1} - 6 \text{ GeV}, \tag{18}$$

can be placed on the allowed region of Eq.(8).

Note that a lower limit of 67 GeV for $\tilde{t}_1$ has been obtained from the direct search for $\tilde{t}_1 \to c\tilde{\chi}_1^0$ under the restriction $M_{\tilde{t}_1} - M_{\tilde{\chi}_1^0} > 10$ GeV [17]. Our results in Fig.2 show that under the condition $M_{\tilde{t}_1} - M_{\tilde{\chi}_1^0} \geq 6$ GeV, Run 2 can explore the entire presently allowed parameter space. In order words, if this final state is not seen at Run 2, only a very small part of parameter space will be allowed, which satisfies Eq.(6) or Eq.(7) plus $M_{\tilde{\chi}_1^0} > M_{\tilde{t}_1} - 6$ GeV.

We conclude by noting that the more precise $t\bar{t}$ cross section measured at Run 2 will further strengthen the upper bound for $B(t \to \tilde{t}_1 \tilde{\chi}_1^0)$ given in Eq.(10). However, as our results show, is it also possible to discover the decay $t \to \tilde{t}_1 \tilde{\chi}_1^0$ in Run 2 for values of $B(t \to \tilde{t}_1 \tilde{\chi}_1^0)$ much smaller than the upper limits obtained by measuring the $t\bar{t}$ cross section. In any event, further constraints on the parameter space can be made through searching for this final state.

## 4. Summary

In the framework of the MSSM with the lightest neutralino being the LSP, we first determined the allowed range for the branching fraction $B(t \to \tilde{t}_1 \tilde{\chi}_1^0)$ in the present allowed region of parameter space and found a lower bound of 0.07. Then we investigated the possibility of observing $t \to \tilde{t}_1 \tilde{\chi}_1^0$ at the Tevatron by searching for the final state $t\bar{t} \to Wbc\tilde{\chi}_1^0\tilde{\chi}_1^0$. We found that:

(a) This final state is unobservable at Run 1 ;

(b) Run 2 can either discover this new decay mode or place the additional constraint $M_{\tilde{\chi}_1^0} > M_{\tilde{t}_1} - 5$ GeV if $S \geq 3\sqrt{B+S}$ is required for discovery of the signal, or $M_{\tilde{\chi}_1^0} > M_{\tilde{t}_1} - 6$ GeV if we require $S \geq 5\sqrt{B}$.

In our analysis, we neglected the possibility of the enhancement of the top pair production cross section in the MSSM. In particular, the gluino pair production might be significant,[14] and would give rise to a new final state $t\bar{t}\tilde{t}\tilde{t}^*$. This will not affect our conclusion significantly since it will give a final state with more jets than the signal we are considering. With such a mechanism of exotic top pair production, the upper bound on $B(t \to \tilde{t}_1 \tilde{\chi}_1^0)$ can be relaxed up to 50%[14], which will enhance the observability of this new mode at the Tevatron and strengthen our conclusion.

## Acknowledgements

This work was supported in part by the U.S. Department of Energy, Division of High Energy Physics, under Grant Nos. DE-FG02-91-ER4086 DE-FG02-94-ER40817, and DE-FG02-92-ER40730. JMY acknowledges partial support provided by the Henan Distinguished Young Scholars Fund. XZ was supported in part by National Natural Science Foundation of China.

**Figure Captions**

Fig. 1 The value of $B(t \to \tilde{t}_1 \tilde{\chi}_1^0)$ versus $M_{\tilde{\chi}_1^0}$ for the signal to be observable at Run 2 under the criterion $S \geq 3\sqrt{B+S}$. The region above the curve is the observable region.

Fig. 2 Same as Fig. 1, but under the criterion $S \geq 5\sqrt{B}$.

Table 1:
Typical signal and background cross sections in units of fb after various cuts at the Tevatron. The basic cuts are $p_T^{\text{all}} \geq 20$ GeV , $\eta_{\text{all}} \leq 2.5$ and $\Delta R \geq 0.5$, and the transverse mass cut is $m_T \geq 90$ GeV. The signal $t\bar{t} \to Wb\bar{c}\tilde{\chi}_1^0\tilde{\chi}_1^0$ results were calculated by assuming $M_{\tilde{t}_1} = 60$ GeV and $M_{\tilde{\chi}_1^0} = 40$ GeV. We have also everywhere assumed the use of silicon vertex tagging of the $b$-quark jet with 50% efficiency and the probability of 0.4% for a light quark jet to be mis-identified as a $b$-jet. The charge conjugate channels have been included. The numerical results do not include the branching fractions for the top and anti-top decays; the actual cross sections are found by multiplying the given cross sections by the branching fraction factor in the last column, where $x$ stands for $B(t \to \tilde{t}_1\tilde{\chi}_1^0)$.

|  | Run 1 | | Run 2 | | BF factor |
| --- | --- | --- | --- | --- | --- |
|  | basic cuts | basic+$m_T$ cut | basic cuts | basic+$m_T$ cut |  |
| $t\bar{t} \to Wb\bar{c}\tilde{\chi}_1^0\tilde{\chi}_1^0$ | 206 | 114 | 287 | 162 | 2x(1-x) |
| $qb \to q't$ | 79.5 | 2.31 | 116 | 4.96 | $1-x$ |
| $q\bar{q}' \to tb$ | 32.0 | 1.77 | 39.0 | 2.25 | $1-x$ |
| $Wbb$ | 113 | 2.04 | 132 | 2.50 | 1 |
| $Wjj$ | 392 | 2.30 | 505 | 2.88 | 1 |
| $t\bar{t}$ | 5.69 | 2.72 | 7.9 | 3.82 | $(1-x)^2$ |

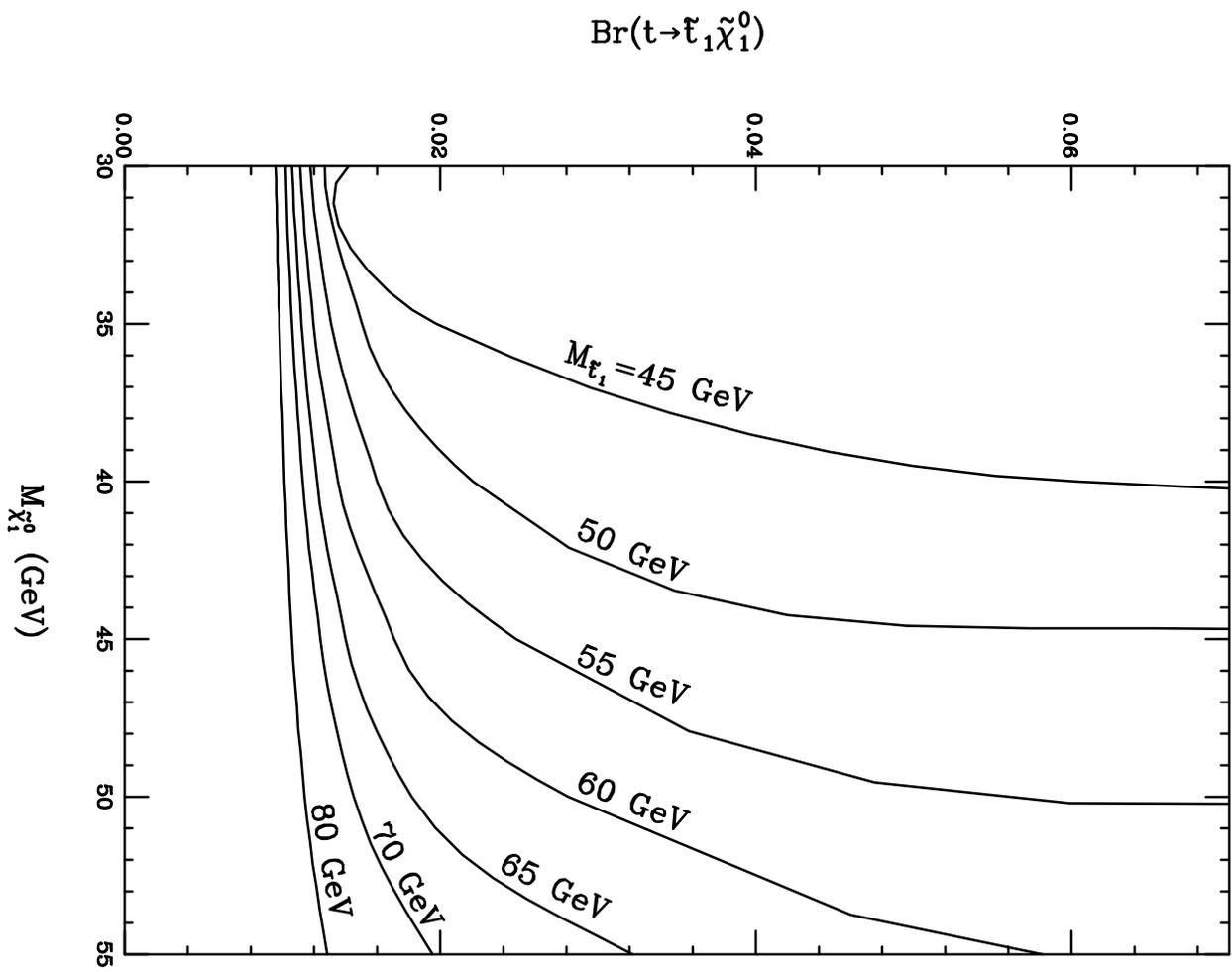

Fig. 1

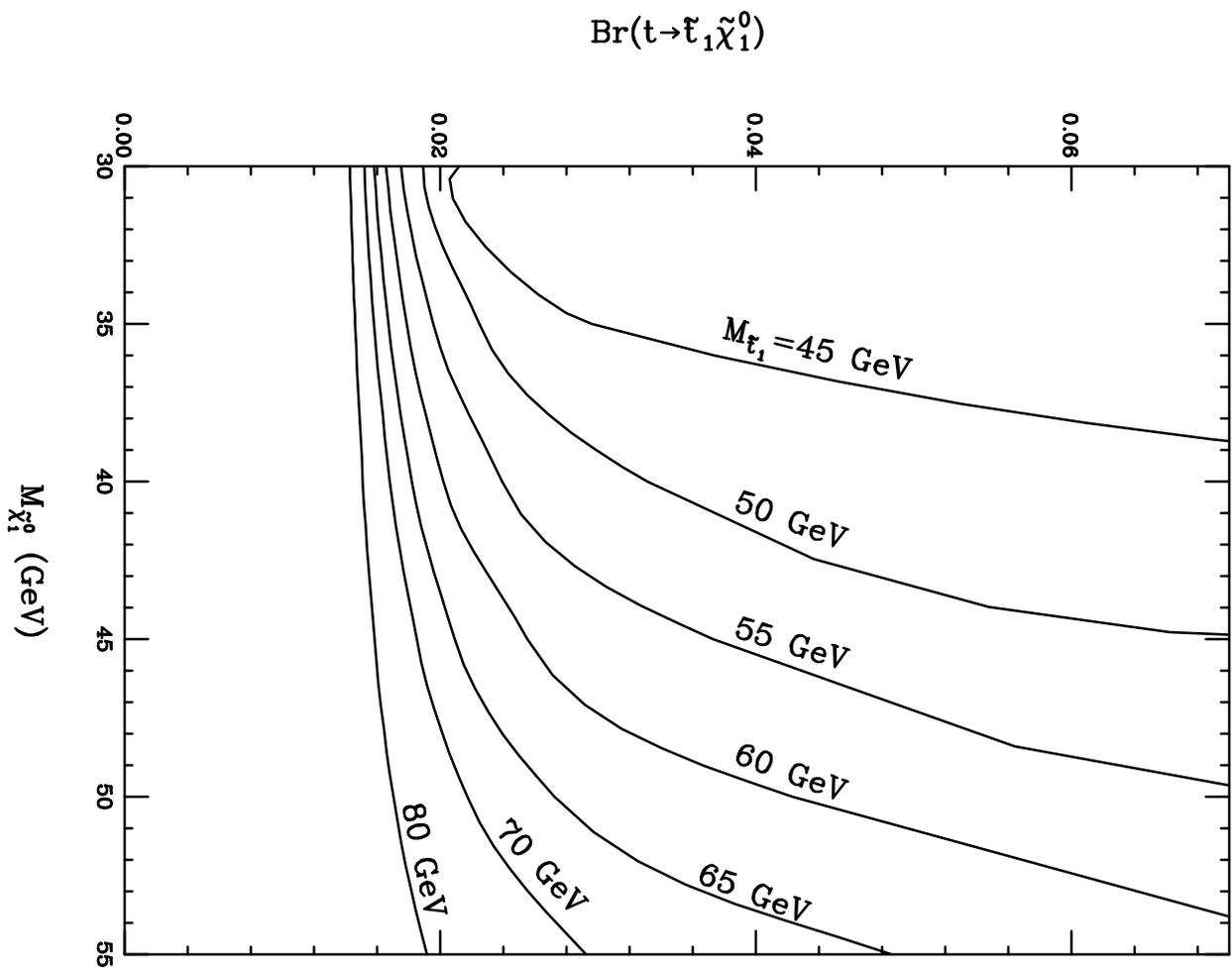

Fig. 2